# Study of the elastocaloric effect and mechanical behavior for the NiTi shape memory alloys[*]

Min Zhou (周敏)[1†], Yushuang Li (李玉霜)[2], Chen Zhang (张晨)[2], and Laifeng Li (李来风)[1†]

[1]*Key Laboratory of Cryogenics, Technical Institute of Physics and Chemistry, Chinese Academy of Sciences, Beijing 100190, China*
[2]*School of Materials Science and Engineering, Beihang University, Beijing 100191, China.*

The NiTi shape memory alloy exhibited excellent superelastic property and elastocaloric effect. Large temperature changes ($\Delta T$) of 30 K upon loading and -19 K upon unloading were obtained at room temperature, which were higher than those of the other NiTi-based materials and among the highest values reported in the elastocaloric materials. The asymmetry of the measured $\Delta T$ values between loading and unloading process was ascribed to the friction dissipation. The large temperature changes originated from the large entropy change during the stress-induced martensite transformation (MT) and the reverse MT. A large coefficient-of-performance of the material ($COP_{mater}$) of 11.7 was obtained at $\varepsilon$=1%, which decreased with increasing the applied strain. These results are very attractive in the present solid-state cooling which is potential to replace the vapor compression refrigeration technologies.

**Keywords:** elastocaloric effect, shape memory alloy, martensitic transformation, entropy change

**PACS:** 65.40.gd, 46.25.Hf, 62.20.fg

## 1. Introduction

Due to the large latent heat associated with the martensitic phase transformation, shape memory alloys were interesting candidates for solid state refrigeration[1]. Compared with the magnetocaloric[2-4] and electrocaloric[5, 6], the elastocaloric cooling attracted much more attention because of its high coefficient of performance (COP) and moderate cost in the past several years [7-11]. The elastocaloric cooling technique has been assessed as the most promising non-vapor compression mechanical refrigeration system for the future by US Department of Energy[12] and opened up new routes for solid-state refrigeration.

Elastocaloric cooling was based on the diffusionless first order phase transformation of the shape memory alloys. The stress-induced martensite transformation led to the heating of the materials. After releasing the stress, the temperature of the materials decreased due to the reverse martensite transformation.

[*] Project supported by the Key Laboratory of Cryogenics, TIPC, CAS (CRYOQN201501 and CRYO201218) and the National Natural Science Foundation of China (Grant No. 51577185, 51377156 and 51408586).
[†] Corresponding author. E-mail: mzhou@mail.ipc.ac.cn (Min Zhou), laifengli@mail.ipc.ac.cn (Laifeng Li)

So, the elastocaloric effect could be regarded as the entropy change under isothermal conditions ($\Delta S_{iso}$) and temperature change under adiabatic conditions ($\Delta T_{adi}$) when a mechanical stress was applied or released in a given material[9]. The coefficient-of-performance (COP) was an important parameter of cooling technology, which described the energy conversion efficiency in the operating temperatures. For the convenience of estimation, the coefficient-of-performance on the material level $COP_{mater}$ (=Q/W) was calculated by assuming that the materials undergo a specific cooling cycle with idea system configuration with full recoverable unloading energy and no auxiliary power consumption [11, 13, 14], where Q was the cooling power, W was the input work.

Compared with other shape memory alloys (FePd[15], CuZnAl[16, 17], NiFe- and NiMn-based SMAs[11, 18-22]), the NiTi-based shape memory alloys [7, 10, 13, 23, 24] showed wonderful potential due to their good elastocaloric effect. In poly-crystalline NiTi wires, Cui et al. reported large temperature changes ($\Delta T$) of 25.5K upon loading and -17K upon unloading for tension cycle. The obtained $COP_{mater}$ values were 3.7 for tension cycle and 11.8 for compression cycle, respectively[13]. Large temperature changes of 21K upon loading and -19K upon unloading (at 342 K) were also reported in the loading-unloading trained NiTi alloys with the applied strain of 6%. Such high temperature changes were contributed to the large entropy changes (35.1 J/kg K for loading and 33.9 J/kg K for unloading process estimated by using the Clausius-Clapeyron equation) [25]. Pataky and Wu et. al then reported the elastocaloric effect of the NiTi single crystals[22, 26]. In Pataky's work, temperature drop of 14.2 K and 13.3 K were observed in the [148] orientation and in the [112] orientation, respectively[22]. In Wu's work, higher temperature drop of 18.2 K was obtained in the [148] orientation[26]. Although the temperature changes of these NiTi single crystals were not higher than the above values of poly-crystal NiTi wires, the estimated entropy changes (maximum $\Delta S$ value of 69.64 J/kg K in [148] orientation) of these NiTi single crystals were much higher than those of the poly-crystal NiTi wires. It is worth noting that the above stress-induced entropy changes were calculated by the Clausius-Clapeyron relationship $\Delta S_\sigma \approx -v_0 \Delta \varepsilon \, d\sigma_t/dT$ [1]. For the NiTi single crystals, the transformation strain $\Delta \varepsilon$ in the above equation was estimated using Lattice Deformation Theory (LDT)[22, 26], which maybe higher than that obtained by the measured stress-strain curves. Besides, the elastocaloric effect of the NiTi (Cu) films was also reported[10, 23, 27].

Compared with single crystals and films, poly-crystals elastocaloric bulks would be suitable for the large-scale application in the solid-state refrigeration. Herein, we studied the elatocaloric effect of the poly-crystalline NiTi bulks. Some key parameters ($\Delta T_{adi}$, $\Delta S_{iso}$, $COP_{mater}$) of elastocaloric effect were also estimated and discussed in a temperature range of 35 K (286-321 K), which was required for almost all practical solid-state cooling applications. The loading-unloading cycles of up to 1000 times were tested under a large strain level of 7%. The present poly-crystalline NiTi SMAs exhibited excellent elastocaloric effect, showing potential prospects in the solid-state refrigeration (or heat-pump) technologies.

## 2. Experimental methods

The starting materials were the poly-crystalline NiTi shape memory alloys. They were annealed with the austenite finish temperature of 283 K ($A_f$). The gauge length of the specimen between the two grips was 60 mm. The uniaxial tensile tests were conducted on a testing machine (20KN, SUNS). The material testing system was equipped with cryogenic furnace. The stress-strain curves at different temperatures were recorded at a low strain rate of $1 \times 10^{-4}$ s$^{-1}$ to ensure isothermal condition. For elastocaloric cooling measurement, the sample was loaded at a much higher strain rate of $5 \times 10^{3}$ s$^{-1}$ (the maximum loading/unloading rate of the testing machine in the strain control mode) to approximately approach the adiabatic condition, and then held for several minutes to make sure that the specimen temperature returned back to the environment temperature. Then, the sample was unloaded quickly. The temperature change during the loading, holding and unloading processes was monitored by a platinum resistance thermometer attached on the middle position of the sample. To improve thermal contact, the platinum resistance thermometer was attached by using silver paint and then was firmly kept in place by means of Teflon tape. The output of the platinum resistance thermometer was read by the Keithley-2000 multimeter at a frequency of 2 Hz. The strain was measured by an electronic extensometer (YYU-10/50, Central Iron and Steel Research Institute at room temperature, and 3542-025M-050-LT, Epsilon at cryogenic temperature). It is worth noting that quicker unloading rate did not result in larger temperature drop in our previous work[28], so the unloading rate of $5 \times 10^{-3}$ s$^{-1}$ was regarded as a reasonable value in this paper.

## 3. Results and discussion

### 3.1. Mechanical behaviors

Fig. 1 (a) shows the stress-strain curves of the NiTi alloys with the applied strain of 7% at various temperatures. The NiTi alloys were tensile tested at approximately isothermal (with low strain rate of $1 \times 10^{-4}$ s$^{-1}$). In the present case, the stress linearly increased with strain at low strain level ($\varepsilon = 1\%$), corresponding to the elastic response of austenite phase. When the critical stress ($\sigma_t$) was reached, a large transformation strain ($\Delta \varepsilon$) occurred at almost constant stress over the transformation plateau. Then the produced martensite phase was elastically deformed. After stress released, the total strain could be fully recovered, indicating complete superelastic deformation in NiTi alloy. It is worth noting that the flat "plateau" appeared in both the loading and unloading processes of the stress-strain curves, typically indicating the positive/reverse stress-induced MT. The "plateau" gradually moves to the higher stress with increasing temperature. As shown in Fig. 1(b), the critical transformation stress ($\sigma_t$) linearly increased with a slope of $d\sigma_t/dT = 8.194$ MPa K$^{-1}$ in the loading and 6.433 MPa K$^{-1}$ in the unloading process. Fig. 1(c) showed the variation of isothermal hysteresis loop area for loading-unloading cycles. Analogous to the behavior of the critical stress, the hysteresis loop area increased with increasing temperature. Both of them showed that more input work was needed to induce the martensite transformation at higher temperatures. It is worth noting that the hysteresis loop area turned to decrease at 321 K, which may be related to the insufficient phase transformation for the applied strain of 7% at higher temperature. Based on the stress-strain curves in Fig. 1(a), the Young's modulus was evaluated and shown in Fig. 1(d). The Young's modulus were strongly temperature dependent and decreased with decreasing temperature, showing softening trend towards the transition temperatures ($A_f$=283 K). Quantitatively, the austenite phase of the NiTi alloy exhibited the softening of approximately 0.94 GPa K$^{-1}$. Analogous results were also reported in other NiTi alloys[29, 30].

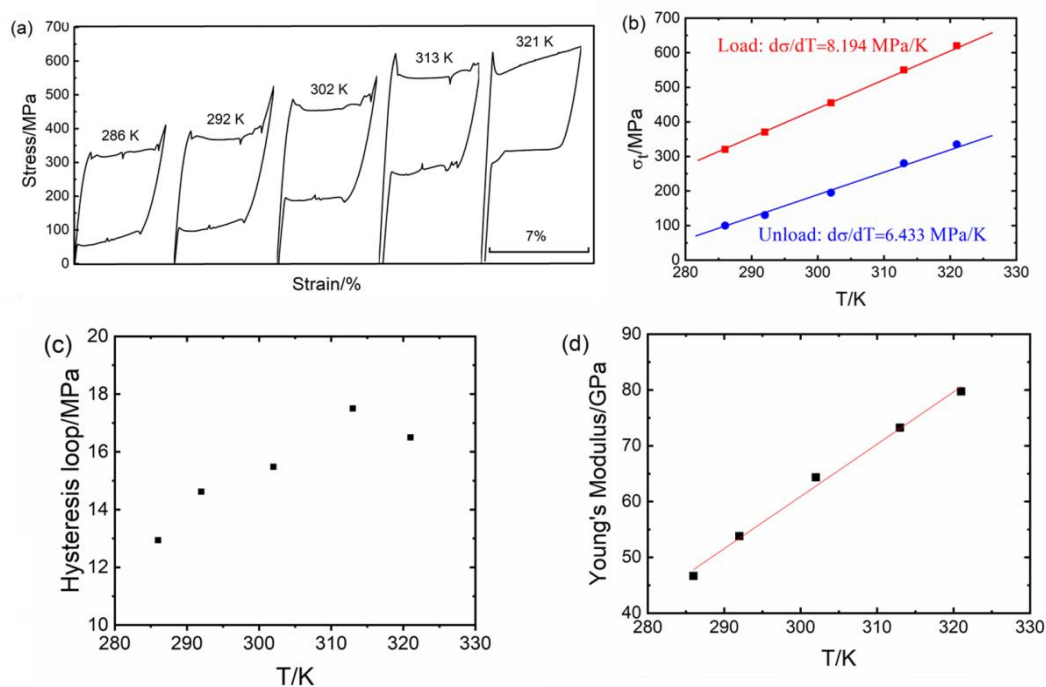

**Fig. 1** (a) Isothermal stress-strain curves of NiTi alloys with the applied strain of 7% at different temperatures. (b) Critical stresses of the stress-induced MT and the reverse MT as a function of temperature. (c) Isothermal hysteresis loop area of the above isothermal stress-strain curves at different temperatures. (d) Young's modulus of NiTi alloys at different temperatures.

## 3.2. Elastocaloric properties

Fig. 2(a) showed the isothermal stress-strain curves (with low strain rate of $1 \times 10^{-4}$ s$^{-1}$) with different strain levels of 1-7% at room temperature. The critical stress ($\sigma_t$) was reached at the applied strain of about 1%. Then the transformation plateau entended with increasing the applied strain. When the applied strain reached to 7%, the stress-induced martensite transformation completed and the produced martensite phase was elastically deformed. After stress released, the total strain could be fully recovered for all the applied strain levels, indicating an ideal superelastic deformation in NiTi alloy. The critical stress reached to about 480 MPa with the applied maximum strain of 2% and then gradually reduced to 420 MPa when the applied maximum strain increased to 7%. The generation of internal stress with increasing strain favored in the formation of stress-induced martensite phase and led to the gradual degradation of the critical stress[25]. We also observed that the area of the stress hysteresis loops between the loading and unloading curves increased with the strain, showing that the fraction of martensite phase increased with the strain.

In order to measure the adiabatic temperature change, an increased strain rate of $5\times10^{-3}$ s$^{-1}$ was used in the tensile test (Fig. 2b) and the temperature changes were shown in Fig. 2c. The temperature change increased correspondingly with the strain, which was attributed to the increase of the martensite volume fraction. When the applied strain increased to 7%, the temperature change ($\Delta T_{loading}$) reached to its maximum value of 30 K upon loading due to the release of heat during the forward MT. During holding, the temperature change value recovered to the room temperature as a result of heat exchange between the sample and environment, and then decreased to -19 K upon unloading ($\Delta T_{unload}$) owing to the absorption of heat during reverse MT. At last, the temperature change value recovered again to the room temperature during further holding due to the heat exchange.

The $\Delta T_{loading}$ values were not equal to the $\Delta T_{unloading}$ values at each strain level (1-7%) (Fig. 2c), indicting the elastocaloric irreversibility ($\Delta T_{irr}=\Delta T_{load}-\Delta T_{unload}\neq0$) between the forward and reverse MTs. Analogous elastocaloric irreversibility were also reported in other NiTi alloys[28]. We also calculated the elastocaloric irreversibilities derived from the friction dissipation ($\Delta T_{fri}$) using the following equation[25]:

$$\Delta T_{fri}=(T\ \Delta S_{fri})/C_p=(T\ Q_{hys}\ )/(C_p\ T)=\oint(\sigma\ d\varepsilon)/(\rho\ C_p) \qquad (1)$$

Where the density ($\rho$) is 6.71 g cm$^{-3}$, the specific heat ($C_p$) is measured as 0.432 J g$^{-1}$ K$^{-1}$ for NiTi alloy. The $Q_{hys}$ value is the stress hysteresis areaof the isothermal tests. The irreversible entropy change ($\Delta S_{fri}$) is obtained from the stress hysteresis area of the stress-strain curves at isothermal condition (Fig. 2a) since the stress hysteresis at adiabatic condition additionally includes the thermodynamic work needed to perform the cooling cycle with self-heating and self-cooling of the material[25]. The obtained $\Delta T_{fri}$ values were plotted in Fig. 2d. They were consistent with the measured $\Delta T_{irr}$ values, which confirmed that the friction dissipation contributed to the elastocaloric irreversibility.

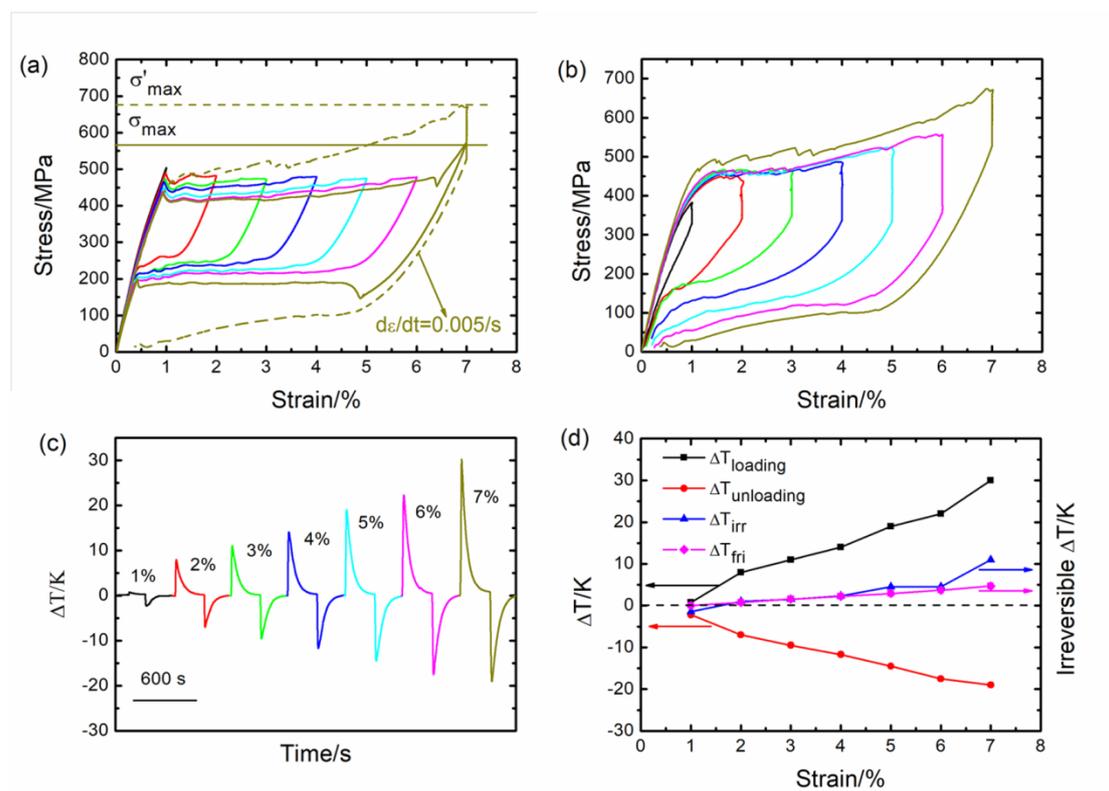

**Fig. 2** (a) Stress-strain curves at different strain levels (ε=1-7%) with approximately isothermal condition (low strain rate of $1×10^{-4}$ s$^{-1}$). (b) Stress-strain curves at different strain levels (ε=1-7%) with approximately adiabatic condition (high strain rate of $5×10^{-3}$/s) at room temperature. The higher strain rate led to both higher critical stresses and larger stress hysteresis. (c) The corresponding ΔT-time profiles at different strain levels (ε=1-7%). (d) The ΔT-strain profiles at different strain levels (ε=1-7%) during the loading and unloading processes. All the above tensile tests were conducted at room temperature.

Fig. 3 showed the measured temperature changes of NiTi alloys in a wider temperature range of 286-321 K, which was required for almost all practical solid-state cooling applications. Large temperature changes of 20~30 K upon loading and -13 ~-19 K upon unloading were observed with the applied strain of 7% in the measured temperature range. The maximum temperature changes (ΔT$_{max}$) were 30 K upon loading and -19 K upon unloading at room temperature, which were even higher than the measured ΔT values of some other reported NiTi alloys (such as the poly-crystalline NiTi wires[13, 28], the trained poly-crystalline NiTi wires[25], the NiTi single crystal[22, 26] and the NiTi thin film[27]).

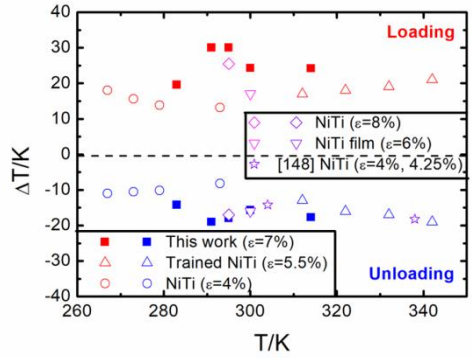

**Fig. 3** The measured temperature changes of NiTi alloys upon loading and unloading process with the applied strain of 7%. The open triangles corresponded to the measured ΔT values of the trained poly-crystalline NiTi alloy with the applied strain of 5.5%[25]. The open cycles corresponded to the measured ΔT values of the poly-crystalline $Ni_{50.8}Ti_{49.2}$ alloy with the applied strain of 4%[28]. The open diamonds corresponded to the measured ΔT values of the poly-crystalline NiTi wires with the applied strain of 8%[13]. The open down-triangles corresponded to the measured ΔT values of the NiTi films with the applied strain of 6%[27]. The open stars corresponded to the measured ΔT values in [148] orientation of the NiTi single crystals with the applied strain of 4%[26] and 4.25%[22].

Based on the Clausius-Clapeyron equation, the stress-induced entropy change ($\Delta S_\sigma$) could be calculated as $\Delta S_\sigma \approx \Delta S_{iso} = -v_0 \Delta \varepsilon \; d\sigma_t/dT$ [1]. Where T was the ambient temperature in Kelvin, $C_p$ was the specific heat (432 J kg$^{-1}$ K$^{-1}$), $v_0$ was the specific volume of $1.48\times10^{-4}$ m$^3$ kg$^{-1}$), $\Delta\varepsilon$ was the transformation strain, and $d\sigma_t/dT$ was the critical transformation stress dependence on temperatures (shown in Fig. 1b). As a result, large $\Delta S_\sigma$ values of -60.6~-75.8 J kg$^{-1}$ K$^{-1}$ for loading and 45.7~49.5 J kg$^{-1}$ K$^{-1}$ for unloading process were obtained in the measured temperature range of 286-321 K (Fig. 4b), which were higher than those of some other poly-crystalline NiTi alloys[25, 28]. So the above high temperature changes could be attributed to these large entropy changes. In the NiTi single crystals, a larger theoretical $\Delta S_\sigma$ value of 69.64 J kg$^{-1}$ K$^{-1}$ was reported in the [148] orientation, in which the transformation strain ($\Delta\varepsilon_{cal}$) was calculated using Lattice Deformation Theory (LDT)[22]. The $\Delta\varepsilon_{cal}$ value is the potential attainable transformation strain. So the measured temperature drop (-14 K) is not higher than our results (shown in Fig. 3).

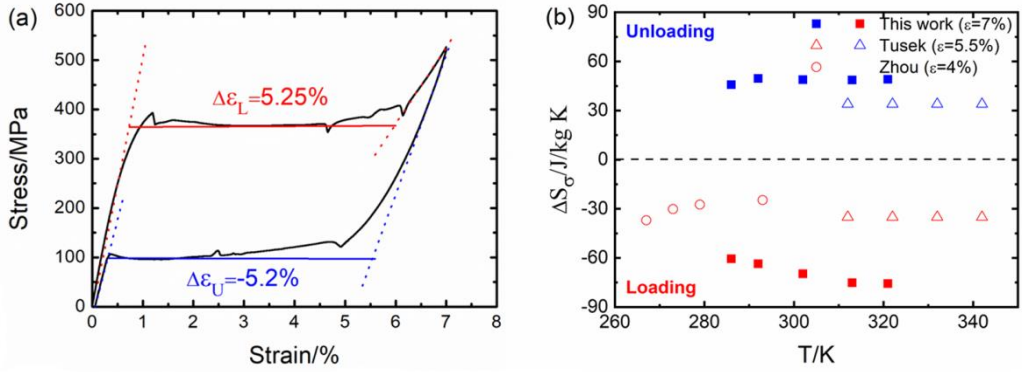

**Fig. 4** (a) The transformation strains ($\Delta\varepsilon$) during the stress-induced MT and the reverse MT at room temperature. (b) The stress-induced entropy change ($\Delta S_\sigma$) of NiTi alloys. The open cycles corresponded to the data of the poly-crystalline $Ni_{50.8}Ti_{49.2}$ alloy[28]. The open triangles corresponded to the data of the trained poly-crystalline NiTi alloy[25].

As mentioned above, $COP_{mater}$ is expressed by the ration of cooling power (Q) to input work (W) ($COP_{mater} = Q/W = \Delta T_{adi} C_p \rho/W$)[11, 13]. Where, $\Delta T_{adi}$ is the theoretical temperature change in the adiabatic condition, which is a potential attainable value. In the following work, the $COP_{mater}$ value was approximately estimated ($COP_{mater} \approx \Delta T_{mea} C_p \rho/W$) by using the measured temperature change ($\Delta T_{mea}$). Based on the adiabatic stress-strain curves at room temperature (Fig. 2b), the input work (W) in the unloading process was calculated by integrating the area enclosed by the loading and the unloading curves[11, 13]. With the density of 6710 kg m$^{-3}$ and heat capacity of 432 J kg$^{-1}$ K$^{-1}$, the $COP_{mater}$ value was estimated at different strain levels ($\varepsilon=1$-7%) and shown in Fig. 5. The $COP_{mater}$ values showed an inverse dependency on the applied strain levels. When the applied strain level was $\varepsilon=1$%, the obtained $COP_{mater}$ value was 11.7. With increasing the applied strain level, the $COP_{mater}$ values decreased. Smaller strain amplitudes should be applied to increase the $COP_{mater}$ values. However, larger strain should be loaded up to obtain high temperature change. So, moderate strain level would be applied in the actual cooling application.

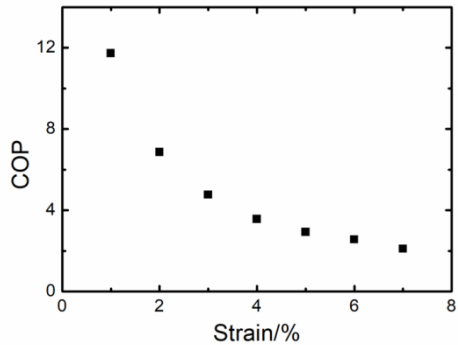

**Fig. 5** The $COP_{mater}$ values of the NiTi alloys at different strain levels ($\varepsilon=1$-7%) during the unloading process.

Tensile cycling test of 1000 times was performed with a high strain rate ($5\times10^{-3}$ s$^{-1}$) at room temperature. The $\Delta T_{max}$ values of the first cycle were 29 K upon loading and -17 K upon unloading under a large strain level of 7% (Fig. 6a and 6b), which was consistent with the above results of other samples. With increasing cycle numbers, the $\Delta T_{max}$ values gradually decreased and then gradually stabilized at 23 K upon loading and -6 K upon unloading when the cycle number was over 200. The decrease of $\Delta T_{max}$ with increasing cycle number was also observed in some other elastocaloric materials[11, 28], which was attributed to the accumulation of dislocations with increasing cycles. In Fig. 6c and 6d, we observed the decrease of the critical stress of martensitic transformation, the area of hysteresis loop and the transformation strain, which confirmed the accumulation of defects with increasing cycles. The initially formed dislocations and remnant martensite, in turn, suppressed the dislocation generation in the subsequent long-term cycles, which caused to the stabilization of the temperature changes. However, it is still not enough to evaluate the elastocaloric stability (long-term stress cycles on the order of $10^6$ would be needed). In the next work, the microstructural modification by doping (Cu, Co, etc.) or incorporating the ductile second phase into grain boundaries, mechanical training pretreatment of elastocaloric materials are expected to enhance mechanical and elastocaloric stability. Moreover, compression, instead of tension, would restrain the fatigue crack growth and cause longer life span.

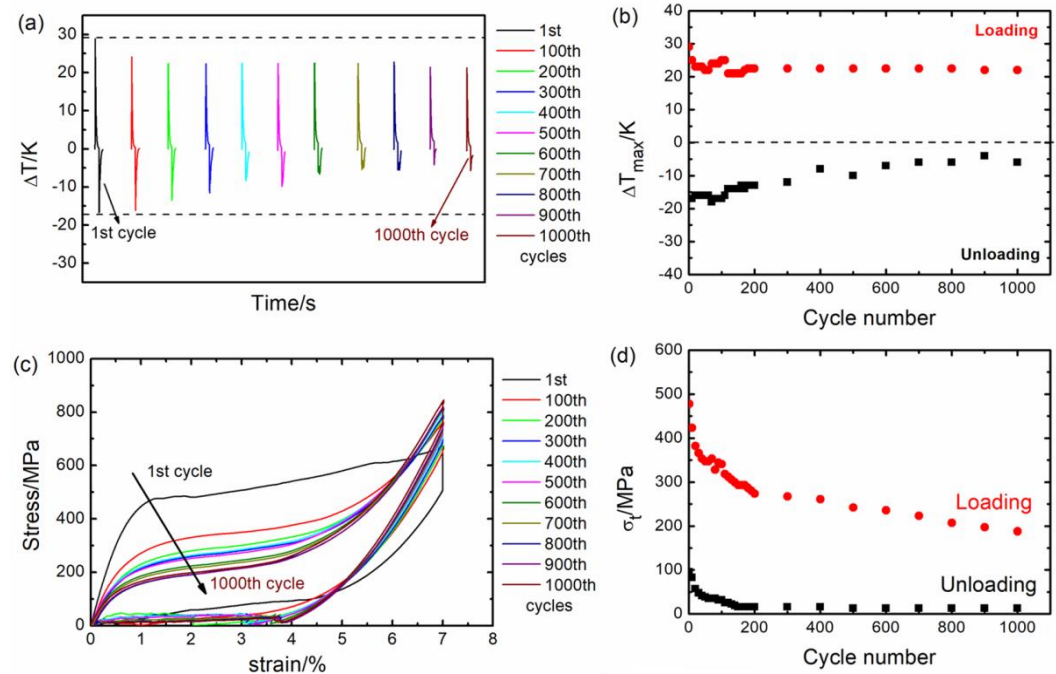

**Fig. 6** (a) The representative temperature variations of the NiTi alloy (with an interval of 100 times, 1st, 100th, 200th, …, and 1000th cycle) during tensile cycling tests with a strain rate of $5\times10^{-3}$ s$^{-1}$ at room temperature. (b) The measured maximum temperature change $\Delta T_{max}$ as a function of the cycle number. (c) The representative stress-strain curves of the NiTi alloys (with an interval of 100 times, the 1st, 100th, 200th, …, and the 1000th cycle). (d) The critical stress of martensitic transformation as a function of the cycle number.

## 4. Conclusions

In summary, the present NiTi alloys not only exhibit excellent shape memory effect and superelasticity but show giant elastocaloric effect. Large temperature changes of 19~30 K upon loading and -13~-19 K upon unloading were measured in a temperature range of 35 K (286-321 K). The maximum $\Delta T$ values ($\Delta T_{max}$) of 30 K upon loading and -19 K upon unloading were measured at room temperature, which were higher than those of the other NiTi-based elastocaloric materials and among the highest values of the elastocaloric materials. The asymmetry of the measured $\Delta T$ values between loading and unloading depended on the applied strain levels and was ascribed to the friction dissipation. The large temperature changes mainly originated from the large entropy change (-60.6~-75.8 J kg$^{-1}$ K$^{-1}$ for loading and 45.7~49.5 J kg$^{-1}$ K$^{-1}$ for unloading process) during the stress-induced MT and the reverse MT. The $\Delta T_{max}$ values decreased in the first 200 mechanical cycles and then gradually stabilized at 23 K upon loading and -6 K upon unloading with increasing tensile cycle numbers up to 1000 cycles, showing the cyclic stability of elastocaloric effect. The accumulation of defects was supposed to contribute to the decrease of $\Delta T_{max}$ values. A high coefficient of performance (COP$_{mater}$) of 11.7 was obtained at $\varepsilon$=1%, which decreased with increasing the applied strain. The present NiTi alloy exhibited larger temperature change ($\Delta T$) and entropy change ($\Delta S_\sigma$) compared with the reported NiTi elastocaloric materials, and showed strong competitive ability in the solid-state refrigeration (or heat-pump) technologies.